\documentclass[epjCONF]{svjour}
\usepackage{graphicx}
\usepackage[varg]{txfonts} 
\usepackage[latin1]{inputenc}
\usepackage{natbib}
\usepackage{sidecap}

\newcommand{\aap}{A\&A}

\session-title{Conference Title, to be filled}
\begin{document}
\title{PLATO: PSF modelling using a microscanning technique}
\author{R-M. Ouazzani\inst{1}\fnmsep\thanks{\email{rhita-maria.ouazzani@phys.au.dk}} \and J.J. Green\inst{2} \and R. Samadi\inst{3}}
\institute{Stellar Astrophysics Centre (SAC), Aarhus University, Ny Munkegade 120 Aarhus C, Denmark, \and Seaview Sensing, Sheffield, South Yorkshire S1 2NS -  United Kingdom \and LESIA, Observatoire de Paris, 5 place Jules Janssen 92195 Meudon cedex, France}
\abstract{
The PLATO space mission is designed to detect telluric planets in the habitable zone of solar type stars, and simultaneously characterise the host star using ultra high precision photometry.
The photometry will be performed on board using weighted masks. However, to reach the required precision, corrections will have to be performed by the ground segment and will rely on precise knowledge of the instrument PSF (Point Spread Function).
We here propose to model the PSF using a microscanning method. 
} 
\maketitle
\section{The Microscanning Method}
\label{sec:1}

%

\subsection{Super-resolved problem}
\label{sec:2}
PLATO will be equipped with 34 12cm aperture telescopes, each one of which composed of 4 CCDs of 4510$\times$4510 pixels (15 arcsec), to cover about 50$\%$ of the sky. The telescopes were designed to be almost focused, except for a slight misalignment of the CCD plane with respect of the focal plane. Therefore, 90$\%$ of the light flux from a star would be collected in a 2$\times$2 pixels square surface. This resolution is insufficient when very precise measurements are desired, which require fine corrections of the jitter (1/3000 CCD pixels) for instance. In this case, the PSF must be modeled with subpixel accuracy. Building a super-resolved PSF from pixel-resolved ones is possible, in practice, through the exploitation of subpixel motion, which, in the case of PLATO, is imposed to the satellite platform during the calibration phases. Consider the reconstruction of a single high-resolution image $X$, given a set of $N$ low-resolution images acquired with different subpixel displacements of the camera. Each acquired image $Y_k$ may be described as: $Y_k \, = \, A_k X$ with $(k\, =\, 1, ...,N)$. The high-resolution image can be retrieved by mean of an inverse method \citep{Park2003}. Here the Landweber iterative method is considered (successfully applied for CoRoT, Pinheiro da Silva et al. 2006). The strategy consists in imposing a slight motion in order to get slightly different views of the PSFs. Here we investigate a microscanning pattern which is an Archimedian spiral, covering a pixel square surface in 256 positions.
\subsection{The pointing strategies}
{\bf Option A:} The motion in spiral is achieved in 256 positions in a discrete manner, assuming: 25 seconds of observation, 25 seconds to move to the next position, 25 seconds to allow the attitude system to engage, in total: 5 hours a 20 minutes. The position is : {\bf A1}: given by the fast telescopes and averaged over 25 seconds; {\bf A2}: directly given by the fast telescopes. Targets are regularly distributed.\\
{\bf Option B:} The strategy is the same as in {\bf A} except here nine targets are used to invert the PSF. \\
{\bf Option C:} Continuous motion along the spiral, and the pointing position is known every 2.5 seconds (fast cameras), the imagettes are integrated every 25 seconds by the normal cameras. The position knowledge is accounted for in the inversion algorithm. The procedure takes 1 hour and 45 minutes.

\section{Tests and performances}
The synthetic data have been simulated using the PLATO Image Simulator (PIS) developed in LESIA, following previous works such as \cite{Marcos-Arenal2014,Arentoft2004}. It models different sources of noise, such as: the satellite high frequency jitter (blurring), the read out smearing, the sky background, the photon noise, the read out noise, electronic noise sources and saturation effects. The simulations have been performed for targets of magnitude $mV=10$. 
We evaluate the quality of the high-resolution PSFs using the 1-norm errors against the original PSF: $L_1 \, = \, \int \, \mid \Delta(x,y) \mid \, dx \, dy$.

\begin{figure}
  \begin{center}
    \begin{minipage}[t]{0.8\linewidth}
      {\vspace*{-1.2cm}\includegraphics[width= 1 \linewidth]{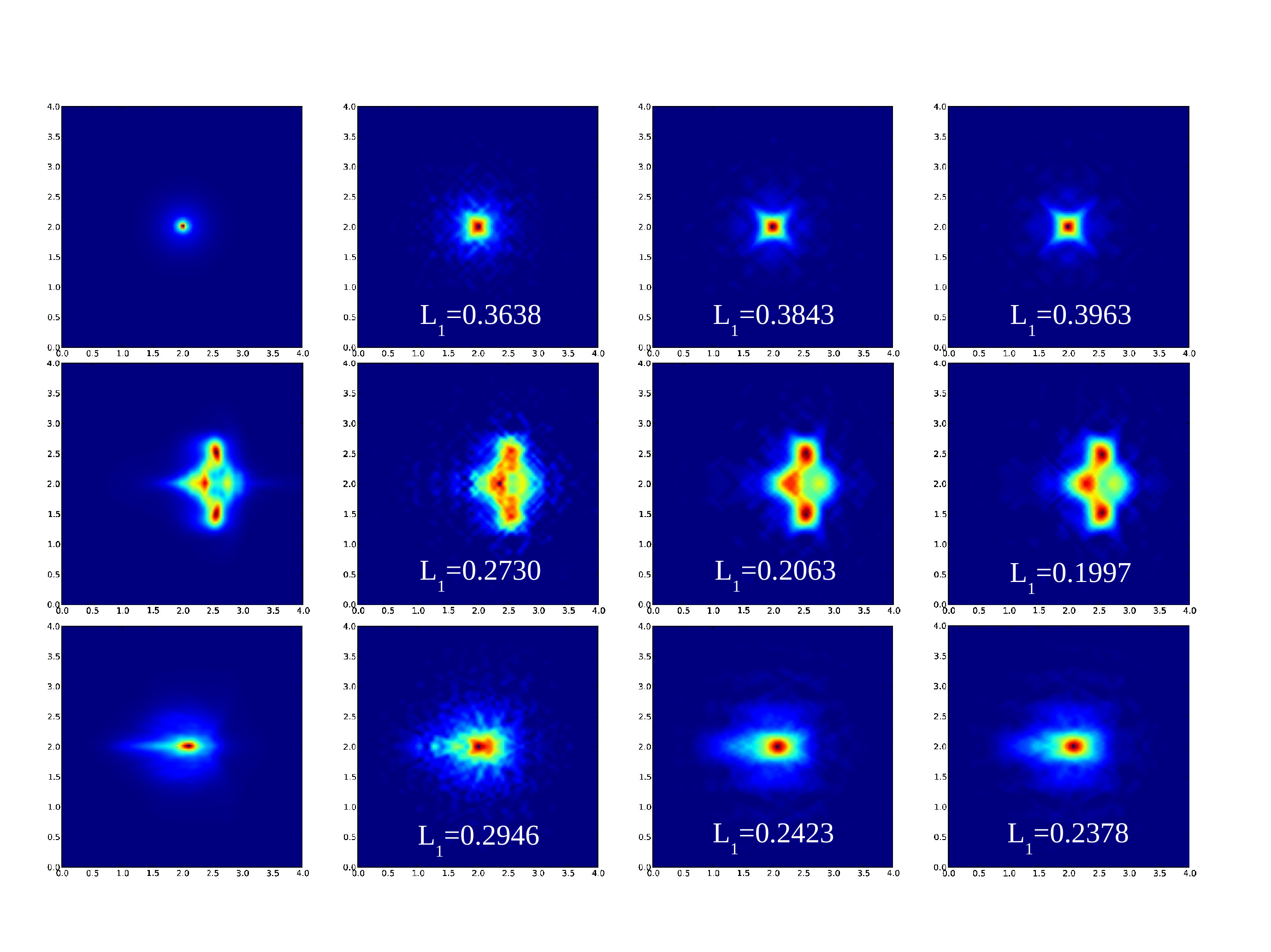} }
    \end{minipage}\hfill
    \begin{minipage}[t]{0.2\linewidth}
      \vspace*{0.2cm}\caption{\label{fig} PSFs for discretization on 32 $\times$ 32 grid inverted from 256 imagettes from archimedian spiral microscanning. {\bf a)} From left to right: for radius and azimuth on the CCD: (1$^{\rm o}$;45$^{\rm o}$), (14$^{\rm o}$;39$^{\rm o}$), (20$^{\rm o}$;28$^{\rm o}$). From top to bottom: theoretical PSF, PSF according for option A1, PSF for option A2, PSF for option C. }
    \end{minipage}
  \end{center}
\end{figure}

The results are presented Fig. \ref{fig}. For the step and stare strategy, the knowledge of the position at the shortest cadence improves the resolution of the PSF's finest structure and gives more precise PSFs. Both A$_2$ and C pointings provide very close values of 1-norms, the differences may not be statistically significant. The real advantage of option C is the time span of the microscanning, which would require much less time during the  calibration phase of the instrument.
Concerning the strategy B, it seems to get better results than the single target ones closer to the optical centre, and similar results further away. The purpose of such a pointing is that taking different targets nearby is equivalent to take different views of the same PSF, as long as the CCDs non-uniformity remains reasonable in that neighborhood. But this strategy doesn't improve drastically the quality of the inverted PSF, probably because an angular separation of $\pm$2$^{\rm o}$ is not sufficient to have significantly different views, and the information given by the different targets are somehow redundant. Larger angular separation between the targets should be investigated further in the future.

However, all these results correspond to ideal cases, more realistic simulations must be undertaken, in particular, the impact of the error on the star displacements must be assessed in the future.

\end{document}